\documentclass[preprint,review,12pt]{elsarticle}

\usepackage{fullpage}

\usepackage{graphicx}

\usepackage{amssymb}
\usepackage{amsmath}
\usepackage{natbib}

 \usepackage{lineno}

\biboptions{square}

\journal{Advances in Water Resources}

\begin{document}

\begin{frontmatter}


\title{Saturated-Unsaturated Flow in a Compressible Leaky-Unconfined Aquifer}


\author[label1]{Phoolendra K. Mishra}
\author[label1]{Velimir V. Vesselinov}
\author[label2]{Kristopher L. Kuhlman}

\address[label1]{Computational Earth Sciences Group, Los Alamos National Laboratory, MS T003, Los Alamos, NM 87545, USA}
\address[label2]{Repository Performance Department, Sandia National Laboratories, 4100 National Parks Highway, Carlsbad, NM 88220, USA}
\begin{abstract}
An analytical solution is developed for three-dimensional flow towards a partially penetrating large-diameter well in an unconfined aquifer bounded below by a leaky aquitard of finite or semi-infinite extent. The analytical solution is derived using Laplace and Hankel transforms, then inverted numerically. Existing solutions for flow in leaky unconfined aquifers neglect the unsaturated zone following an assumption of instantaneous drainage due to Neuman.  We extend the theory of leakage in unconfined aquifers by (1) including water flow and storage in the unsaturated zone above the water table, and (2) allowing the finite-diameter pumping well to partially penetrate the aquifer. The investigation of model-predicted results shows that aquitard leakage leads to significant departure from the unconfined solution without leakage. The investigation of dimensionless time-drawdown relationships shows that the aquitard drawdown also depends on unsaturated zone properties and the pumping-well wellbore storage effects.
\end{abstract}


\end{frontmatter}

\linenumbers

\section{Introduction}
The assumption that the water flow and storage in the unsaturated zone is insignificant for unconfined aquifer tests was first questioned by \citet{nwankwor1984} and later by \citet{akindunni1992} based upon analysis of data collected during pumping tests in Borden, Ontario Canada. Analyzing the collected tensiometer data and soil moisture measurements, the authors concluded that the proper inclusion of unsaturated zone in analytical models used for pumping test analysis would lead to improved estimates of aquifer specific yield. Several analytical solutions were developed that account for the unsaturated zone flow to a pumping well in an unconfined aquifer, taking into account the unsaturated zone \citep{tartakovsky2007,mathias2006,mishra_neuman2010}. These models consider the unsaturated zone effects by coupling the governing flow equations at the water table; the saturated zone governed by the diffusion equation and the vadose zone governed by the linearized unsaturated zone Richards’ equation, using the linearization of \citet{kroszynski1975}. These models considered the limiting case where the pumping well has zero radius. For detailed discussion regarding the fundamental differences between these three models readers are directed to \citet{mishra_neuman2010}.

Drawdown due to pumping a large-diameter (e.g., water supply) well in an unconfined aquifer is affected by wellbore storage \citep{papadopulos1967}. \citet{narasimhan1993} used a numerical model to demonstrate that early time drawdown in an unconfined aquifer tends to be dominated by wellbore storage effects. \citet{mishra_neuman2011} developed an analytical unconfined solution, which considers both pumping-well wellbore storage capacity, and three-dimensional axi-symmetrical unsaturated zone flow. They represented unsaturated zone constitutive properties using exponential models, which result in governing equations that are mathematically tractable, while being sufficiently flexible to be fit to other widely used constitutive models \citep{gardner1958, russo1988, brooks1964, van1980}. However, \citet{mishra_neuman2011} considered the unconfined aquifer to be resting on an impermeable boundary and therefore did not account for the potential effects of leakage from an underlying formation (e.g., an aquitard or fractured bedrock).
The classical theory of leakage for confined aquifers was originally developed by \citet{hantush1955} assuming steady-state vertical flow in overlying and underlying aquitards and horizontal flow in the pumped aquifer. \citet{hantush1960} later modified the theory of confined leaky aquifers to include transient vertical aquitard flow, giving asymptotic expressions for early and late times. \citet{neuman1969a, neuman1969b} developed a more complete analytical solution for the more general multiple aquifer flow problem, but did not consider general three-dimensional aquitard flow.

\citet{yatov1968} first investigated the effect of leakage from underlying strata on unconfined aquifer flow. He use the model of  \citet{boulton1954}  to account for the water table and considered only vertical flow in aquitard. \citet{ehlig1976} investigated leaky-unconfined flow through a finite-difference simulation, which coupled the \citet{boulton1954} and \citet{hantush1955} models to account for leakage across the aquifer-aquitard boundary. \citet{zlotnik_zhan2005} developed an analytical solution for the flow towards a fully penetrating zero-radius well in a coupled unconfined aquifer–aquitard system where both the unsaturated zone and the horizontal aquitard flow are neglected. Both \citet{zhan_bian2006} and \citet{zlotnik_zhan2005}  developed analytical and semi-analytical solutions for leakage due to pumping, building on the works of \citet{hantush1955} and \citet{butler2003}. Both \citet{zhan_bian2006} and \citet{zlotnik_zhan2005} neglect horizontal flow in aquitards. Purely vertical aquitard flow was justified for limiting aquifer/aquitard hydraulic conductivity contrasts by \citet{neuman1969b}. Both \citet{zhan_bian2006} and \citet{zlotnik_zhan2005} only consider a vertically unbounded aquitard. \citet{malama2007} developed a solution for three-dimensional aquitard flow in a finite thickness aquitard, but considered the zero-radius pumping well to be fully penetrating and ignored the flow in unsaturated zone. Here, we develop a more general leaky-unconfined aquifer solution by considering a partially penetrating large-diameter well and including the effects of unsaturated zone flow following \citet{mishra_neuman2011}. The solution is used to investigate the effect of an aquitard on drawdown in overlying unconfined aquifer.  We conclude by investigating the effects of wellbore storage capacity and the unsaturated zone on drawdown observed in the aquitard.

\section{Leaky-Unconfined Theory}
\subsection{Statement of Problem}
\begin{figure}
\begin{center}
\includegraphics[width = 8 cm]{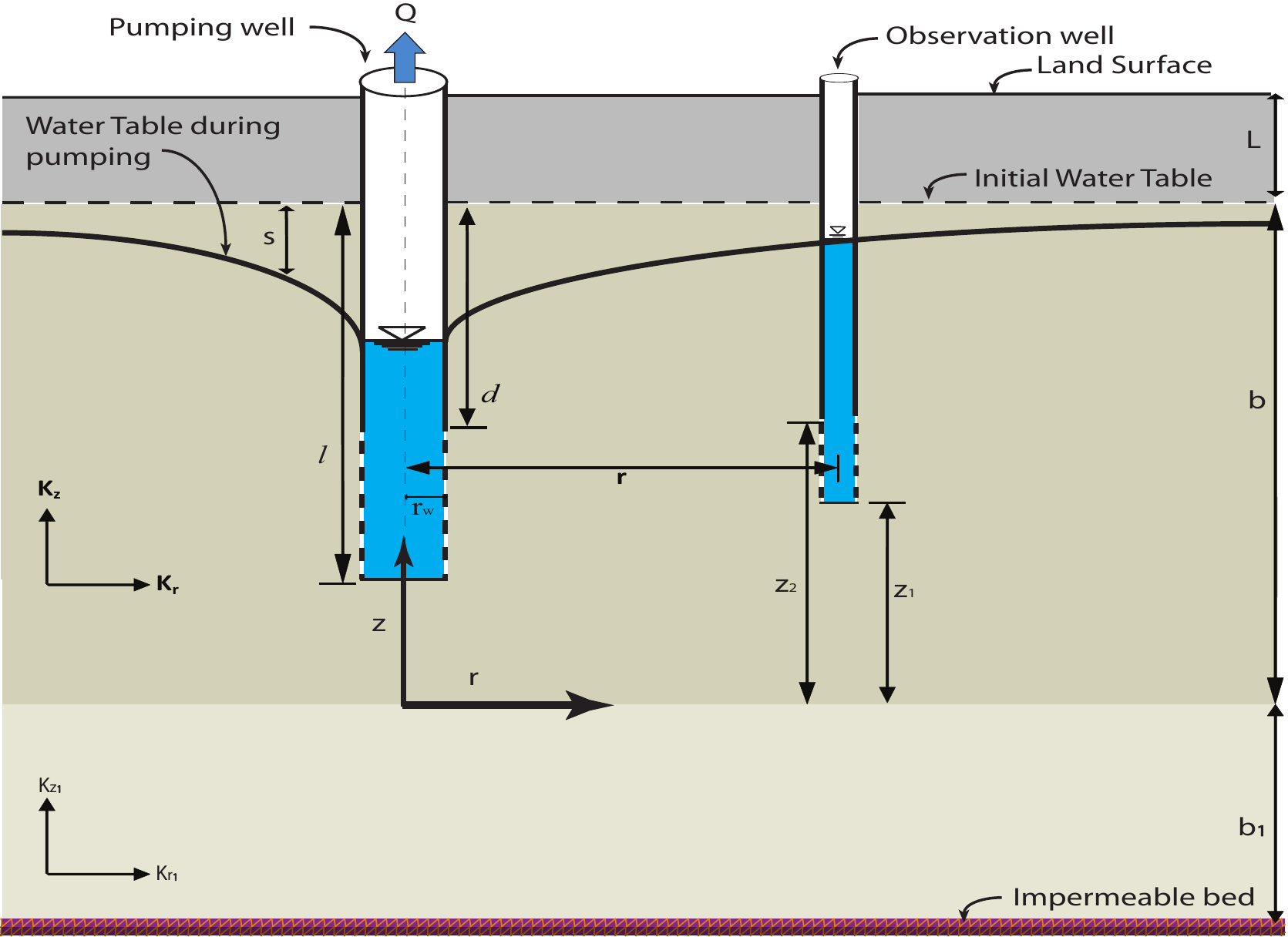}
\caption[Schematic representation of finite radius pumping well in a leaky unconfined aquifer- aquitard system]{Schematic representation of leaky unconfined aquifer-aquitard system geometry with finite radius pumping well\label{fig1}}
\end{center}
\end{figure}

We consider an infinite radial compressible unconfined aquifer above a
finitely thick aquitard (Figure~\ref{fig1}). The aquifer and aquitard are each spatially uniform, homogeneous and anisotropic, with constant specific storage $S_s$ and $S_{s1}$, respectively (a subscript 1 indicates aquitard related properties). The aquifer has a fixed ratio ${{K}_{D}}={{K}_{z}}/{{K}_{r}}$ between vertical and horizontal saturated hydraulic conductivities, ${{K}_{z}}$ and ${{K}_{r}}$, respectively. The aquitard vertical and horizontal hydraulic conductivities are $K_{z1}$  and $K_{r1}$. The aquifer is fully saturated below an initially horizontal water table at elevation $z=b$. The water table is defined as a $\psi = 0$ isobar where $\psi $ is pressure head. A saturated capillary fringe at non-positive pressure ${{\psi }_{a}}\le \psi \le 0$ extends from the water table to the $\psi = \psi_a$ isobar; ${{\psi }_{a}}\le 0$ is the pressure head required for air to enter a saturated medium. The saturated hydraulic system (aquifer and aquitard) is at uniform initial hydraulic head ${{h}_{0}}=b+{{\psi }_{a}}$ before pumping. At time t = 0, pumping begins at a constant volumetric flowrate $Q$ from a well of finite radius ${{r}_{w}}$ and wellbore storage coefficient ${{C}_{w}}$ (volume of water released from storage in the pumping well per unit drawdown in the well casing). The pumping well is completed across the aquifer between depths $l$ and $d$ below the aquifer top. Under these conditions the drawdown $s\left( r,z,t \right)=h\left(r,z,0\right)-h\left( r,z,t \right)$ in the saturated zone is governed by the diffusion equation
\begin{eqnarray}
{{K}_{r}}\frac{1}{r}\frac{\partial }{\partial r}\left( r\frac{\partial s}{\partial r} \right)+{{K}_{z}}\frac{{{\partial }^{2}}s}{\partial {{z}^{2}}}={{S}_{s}}\frac{\partial s}{\partial t}  & r \ge r_{w} & 0\le z<b,
\end{eqnarray}
along with far-field boundary condition
\begin{equation}
s\left( \infty ,z,t \right)=0,
\end{equation}
the no-flow condition at the portion of the well casing that is not open to the aquifer
\begin{equation}
{{\left( r\frac{\partial s}{\partial r} \right)}_{r={{r}_{w}}}}=0 \quad 0\le z\le b-l \quad b-d\le z\le b,
\end{equation}
and the wellbore storage mass-balance expression
\begin{equation}
2\pi {{K}_{r}}\left( l-d \right){{\left( r\frac{\partial s}{\partial r} \right)}_{r={{r}_{w}}}}-{{C}_{w}}{{\left( \frac{\partial s}{\partial t} \right)}_{r={{r}_{w}}}}=-Q  \quad b-l\le z\le b-d.
\end{equation}

Flux is assumed constant across the well screen (see \citet{zhan2002} for a discussion of this assumption's validity). The corresponding linearized unsaturated flow equations \citep{mishra_neuman2010} are

\begin{eqnarray}\label{unsat_flow}
{{K}_{r}}{{k}_{0}}\left( z \right)\frac{1}{r}\frac{\partial }{\partial r}\left( r\frac{\partial \sigma }{\partial r} \right)+{{K}_{z}}\frac{\partial }{\partial z}\left( {{k}_{0}}\left( z \right)\frac{\partial \sigma }{\partial z} \right)={{C}_{0}}\left( z \right)\frac{\partial \sigma }{\partial t}   \\
\quad r\ge {{r}_{w}}  \quad b<z<b+L \nonumber
\end{eqnarray}
where $\sigma(r,z,t)$ is drawdown in the unsaturated zone,  $k_0(z)$ is relative permeability and $C_0(z)$  is moisture capacity (slope of the curve representing water saturation as a function of pressure head) functions with the functional dependence limitations on the respective constitutive models
\begin{equation}
{{k}_{0}}\left( z \right)=k\left( {{\theta }_{0}} \right), \quad {{C}_{0}}\left( z \right)=C\left( {{\theta }_{0}} \right)
\end{equation}
where $\theta_0$ is the initial volumetric moisture content. Equation \eqref{unsat_flow} depends on the initial condition
\begin{equation}
\sigma (r,z,0)= 0,
\end{equation}
the far-field boundary condition
\begin{equation}
\sigma \left( \infty ,z,t \right)=0
\end{equation}
the no-flow condition at the ground surface
\begin{equation}
\left. \frac{\partial \sigma }{\partial z}\right|_{z=b+L} =0 \quad    r\ge {{r}_{w}}
\end{equation}
and the no-flow condition at the well casing
\begin{equation}
{\left( r\frac{\partial \sigma }{\partial r} \right)}_{r=r_w}=0 \quad   b<z<b+L.
\end{equation}
The interface conditions providing continuity across the water table are
\begin{equation}
s-\sigma =0            \quad            r\ge r_w              \quad             z=b,
\end{equation}

\begin{equation}
\frac{\partial s}{\partial z}-\frac{\partial \sigma }{\partial z}=0 \quad r\ge {{r}_{w}} \quad z=b.
\end{equation}

The aquitard drawdown $s_1\left( r,z,t \right)$ is governed by
\begin{equation}
{{K}_{r1}}\frac{1}{r}\frac{\partial }{\partial r}\left( r\frac{\partial {{s}_{1}}}{\partial r} \right)+{{K}_{z1}}\frac{{{\partial }^{2}}{{s}_{1}}}{\partial {{z}^{2}}}={{S}_{s1}}\frac{\partial {{s}_{1}}}{\partial t} \quad                    r\ge 0 \quad               -b_1\le z<0.
\end{equation}
Additionally, aquitard flow satisfies no-flow conditions at the bottom and center of the flow system
\begin{equation}
\underset{r\to 0}{\mathop{\lim }}\,\left( r\frac{\partial {{s}_{1}}}{\partial r} \right)={{\left. \frac{\partial {{s}_{1}}}{\partial z} \right|}_{z=-{{b}_{1}}}}=0 .
\end{equation}

The interface condition across the aquifer-aquitard boundary are
\begin{equation}
s-s_1=0 \quad r\ge r_w \quad z=0
\end{equation}
and
\begin{equation}
K_z\frac{\partial s}{\partial z}=K_{z1}\frac{\partial s_1}{\partial z} \quad r\ge r_w \quad z=0 .
\end{equation}

Like \citet{mishra_neuman2010}, we represent the aquifer moisture retention curve using an exponential function
\begin{equation}\label{eff_sat}
{{S}_{e}}=\frac{\theta \left( \psi  \right)-{{\theta }_{r}}}{{{S}_{y}}}={{e}^{{{a}_{c}}\left( \psi -{{\psi }_{a}} \right)}} \quad {{a}_{c}}\ge 0 \quad \psi_a \ge 0
\end{equation}
where $\theta_r$ is residual volumetric water content, $S_y=\theta_s-\theta_r$ is drainable porosity or specific yield and $S_e$ is effective saturation. We also adopt the exponential relative hydraulic conductivity model \citep{gardner1958},
\begin{equation}
k(\psi )=\left\{ \begin{matrix}
   {{e}^{{{a}_{k}}\left( \psi -{{\psi }_{k}} \right)}}  \\
   1  \\
\end{matrix} \right.\quad \quad \quad \begin{matrix}
   \psi \le {{\psi }_{k}}  \\
   \psi >{{\psi }_{k}}  \\
\end{matrix}           \quad {{a}_{k}}\ge 0 \quad \psi_k \ge 0,
\end{equation}
with parameters ${{a}_{k}}$ and ${{\psi }_{k}}$ that generally differ from ${{a}_{c}}$ and ${{\psi }_{a}}$ in \eqref{eff_sat}. The parameters $a_k$ and $a_c$  represent the exponent in the exponential models for hydraulic conductivity and effective saturation, respectively. The parameter $\psi_k$ represents a pressure head above which relative hydraulic conductivity is effectively equal to unity, which is sometimes but not always equal to the air entry pressure head $\psi_a$.In addition to rendering the resulting equations mathematically tractable, these exponential constitutive models are sufficiently flexible to provide acceptable fits to standard constitutive models such as those mentioned earlier.

\subsection{Point drawdown in saturated and unsaturated zones of aquifer and aquitard}

Following \citet{mishra_neuman2011}, it is shown in Appnendix A that, drawdown in the saturated zone can be expressed as
\begin{equation}
s=s_C + s_U
\end{equation}
where $s_C$ is solution for flow to a partially penetrating well of finite radius in a confined aquifer and $s_U$ is a correction accounting for the underlying aquitard, water table and unsaturated zone effects. The Laplace transformed solution $\bar{s}_C$ is given by \citet{mishra_neuman2011} as
\begin{eqnarray}\label{wellbore_lap_sol}
\bar{s}_{C}\left(r_D,z_D,p_D \right)=\frac{Q}{4\pi Tp_D}\omega_0 \mathrm{K}_0(\phi_0)+\sum \limits _{n=1}^\infty \omega_n \mathrm{K}_0(\phi_n)\cos\left[n\pi (1-z_d)\right] \\ \nonumber
r_d \ge r_w/b
\end{eqnarray}
where 
$\omega_n=\frac{\sin(n\pi l_D)-\sin(n\pi d_D)}{n\pi (l_D-d_D)\Omega(n)}$, 
$\Omega (n)=r_{wD}\phi_0
\mathrm{K}_1(r_{wD}\phi_n)+\frac{C_{wD}}{2(l_D-d_D)}r_{wD}^2\phi_n^2K_0(r_{wD}\phi_n)$, 
$\omega_0=2/\Omega(0)$, 
$r_{wD}=r_w/r$, 
$r_D=r/b$, 
$z_D=z/b$, 
$d_D=d/b$, 
$l_D=l/b$, 
$p_D=pt$, 
$C_{wD}=C_w/(\pi S_s br_w^2)$, 
$t_s=\alpha_st/r^2$, 
$p$ is Laplace parameter (the transform of $t$), 
$\phi_n=\sqrt{p_D/t_s + r_D^2K_Dn^2\pi^2}$, and 
$\mathrm{K}_0$ and $\mathrm{K}_1$ are second-kind modified Bessel functions of orders zero and one.
The Laplace transformed unsaturated zone drawdown $\bar{\sigma}$ is given by \citet{mishra_neuman2011} and is presented in Appendix D for sake of completeness.

The Laplace transformed $\bar{s}_U$ derived in Appendix B is
\begin{equation}\label{su_final}
\bar {s}_U\left(r_D,z_D,p_D\right)=\int_0^\infty \left(\rho_1 e^{\mu z_D}+ \rho_2 e^{-\mu z_D}\right ) \frac{r_D^2K_D}{r^2}yJ_0\left[yK_D^{1/2}r_D \right]\;\mathrm{d}y
\end{equation}
where 
$\rho_1=\frac{\left(\frac{\mu}{qb}+1\right)
  e^{-\mu}\left(\bar{\bar{s}}_c\right)_{z_D=0}-
  \left(\frac{\mu}{q_1b}+1\right)
  e^{-\mu}\left(\bar{\bar{s}}_c\right )_{z_D=1}}{\Delta}$, 
$\rho_2=\frac{\left(\frac{\mu}{qb}-1\right)
e^{-\mu}\left(\bar{\bar {s}}_c\right) _{z_D=0}-
\left(\frac{\mu}{q_1b}-1\right)
e^{-\mu}\left(\bar{\bar {s}}_c\right )_{z_D=1}}{\Delta}$, 
$q_1b=R_{K_z}\mu _1 \tanh \left(\mu_1R_b\right)$, 
$\mu_1^2=\frac{y^2}{R_{K_D}}+
  \frac{p_D}{t_sK_Dr_D^2R_{K_D}R_{\alpha_S}}$, 
$R_{K_D}=K_{D1}/K_D$, 
$R_{K_z}=K_{z1}/K_z$, 
$R_{\alpha_s}=\alpha_{S1}/\alpha_s$, 
$R_b=b_1/b$, 
$\alpha_{s1}=K_{r1}/S_{s1}$, and 
$\Delta =\left(\frac{\mu}{qb}+1\right)\left(\frac{\mu}{q_1b}-1\right)e^{-\mu}-\left(\frac{\mu}{qb}-1\right)\left(\frac{\mu}{q_1b}+1\right)e^{\mu}$.

The Laplace transformed aquitard drawdown derived in Appendix C is
\begin{eqnarray}\label{aquitard_drawdown}
\bar{s}_1(r_D,z_D,p_D)&=&\int_0^\infty
\frac{(\bar{\bar{s_c}})_{z_D=0}+\rho_1+\rho_2}{\cosh(\mu_1b_1/b)} \cosh\left[\mu_1(z_D+R_b)\right]\\ \nonumber
&& \times \frac{r_D^2K_D}{r^2}y \mathrm{J}_0\left[yK_D^{1/2}r_D \right]\; \mathrm{d}y
\end{eqnarray}
where $(\bar{\bar{s_c}})_{z_D=0}$ is the Laplace-Hankel transformed confined aquifer drawdown and is defined in Appendix D.
The time domain equivalents  $s_C$, $s_U$, $s_1$  and $\sigma$  of $\bar{s}_C$ , $\bar{s}_U$, $\bar{s}_1$ and $\bar{\sigma}$  are obtained through numerical Laplace transform inversion using the algorithm of \citet{dehoog1982}.

\subsection{Vertically Averaged Drawdown in Piezometer or Observation Well}
Drawdown in an observation well (Figure~\ref{fig1}) that is completed in the aquifer between elevations ${{z}_{D1}}={{z}_{1}}/b$ and ${{z}_{D2}}={{z}_{2}}/b$ is found by averaging the point drawdown over screen interval,
\begin{equation}
s_{z_{D2}-z_{D1}}(r_D,t_s)=\frac{1}{z_{D2}-z_{D1}}\int_{z_{D1}}^{z_{D2}}s^\star(r_D,z_D,t_s)\;\mathrm{d}z_D
\end{equation}
where $s^\star$ can be either aquifer drawdown $s$, aquitard drawdown $s_1$, or a combination of the two, depending on the observation well screen interval.

\subsection{Delayed Piezometer or Observation Well Response}
When water level is measured in a piezometer or observation well having storage coefficient C the water level observed in the borehole is delayed in time. Following \citet{mishra_neuman2011}, the measured (delayed) drawdown $s_m$  can be expresses in terms of formation drawdown $s$ via
\begin{equation}\label{delayed}
s_m=s\left(1-e^{-t/t_B}\right)
\end{equation}
where $t_B$ is basic (characteristic) monitoring well time lag. The dimensionless equivalent of \eqref{delayed} is
\begin{equation}\label{delayed_dim}
s_{mD}=s_D\left(1-e^{-t_s/t_{Bs}}\right)
\end{equation}
where $t_{Bs}=\alpha_st_B/r^2$, and $r$ is the radial distance to the monitoring location.

\section{Model-predicted drawdown behavior}
We illustrate the impacts of an underlying aquitard on unconfined aquifer drawdown for the case where $K_D=1$, $S_sb/S_y=10^{-3}$, $a_{kD}=a_{cD}=10$, $\psi_{aD}=\psi_{kD}$, $d_D=0$, $C_{wD}=10^3$, $l_D=0.6$ and $r_w/b=0.02$, where $a_{kD}=a_kb$, $a_{cD}=a_cb$, $\psi_{aD}=\psi_a/b$ and  $\psi_{kD}=\psi_k/b$. We also investigate the effects that wellbore storage capacity of the pumping well, the unconfined aquifer, and the unsaturated zone have on aquitard drawdown.

\subsection{Dimensionless unconfined aquifer time-drawdown}
\begin{figure}
\begin{center}
\includegraphics[width = 8 cm]{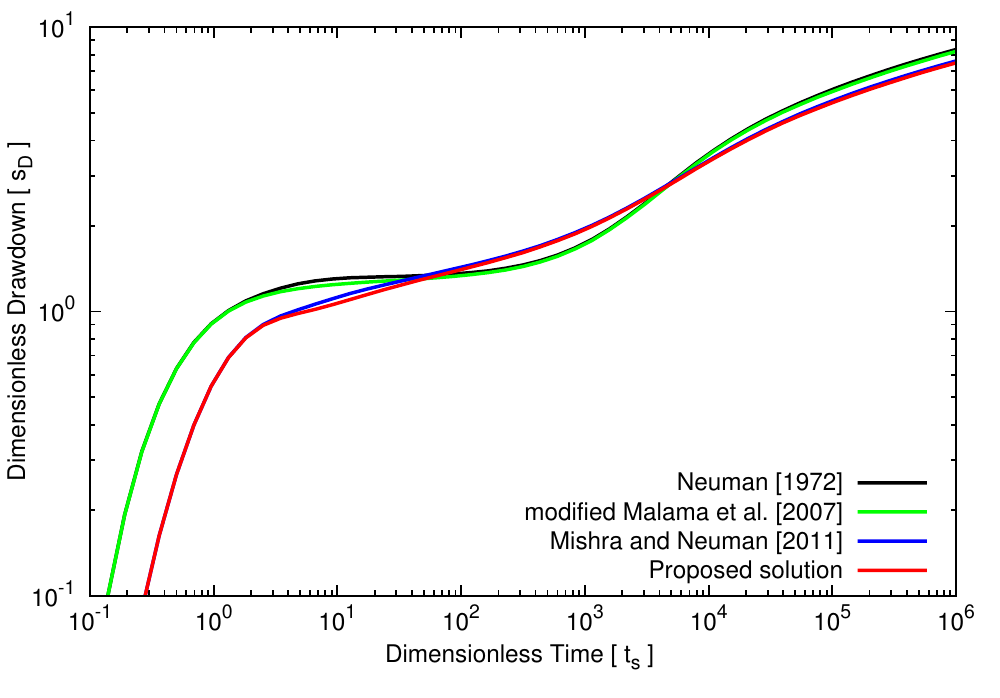}
\includegraphics[width = 8 cm]{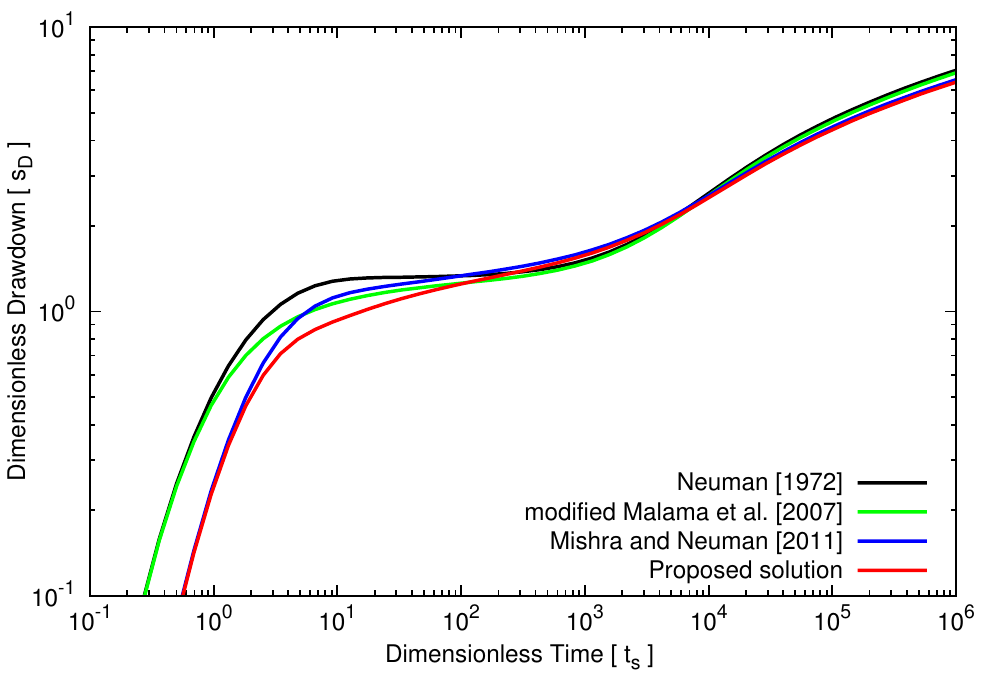}
\caption[Time-drawdown at various depths]{Dimensionless
  leaky-unconfined aquifer drawdown versus dimensionless time at
  $r_D=0.5$  when $K_D=1$, $S_Sb/S_y=10^{-3}$, $a_{kD}=a_{cD}=10$,
  $\psi_{aD}=\psi_{kD}$, $d_D=0$, $l_D=0.6$, $C_{wD}=10^2$,
  $R_{K_r}=R_{K_z}=10^{-2}$, $R_{S_s}=10^{-2}$, $R_b \to \infty$ and
  (a) $z_D=0.75$  (b) $z_D=0.25$. Solutions of
  \citet{mishra_neuman2011}, modified \citet{malama2007}, and
  \citet{neuman1972} are also shown. \label{fig2}}
\end{center}
\end{figure}

We start by considering drawdown at two locations in the unconfined aquifer saturated zone, one location closer to water table ($z_D=0.75$) and the other closer to the aquitard-aquifer boundary ($z_D=0.25$). Figures~\ref{fig2}a and \ref{fig2}b compare variations in dimensionless drawdown $s_D(r_D,z_d,t_s)=(4\pi K_rb/Q)s(r_D,z_D,t_s)$  with dimensionless time at $z_D=0.75$  and $z_D=0.25$  predicted by our proposed solution and the solutions of \citet{mishra_neuman2011}, \citet{neuman1972}, and the modified solution of \citet{malama2007} (modified to include the partially penetrating pumping well effects, as done in \citet{malama2008} for a multi-aquifer system). The solutions of \citet{neuman1972} and \citet{malama2007} do not include wellbore storage effects, and therefore they overestimate drawdown at early time. Both of these solutions also ignore the unsaturated zone above the water table, considering the water table a material boundary \citep{neuman1972}. Our proposed solution follows \citet{mishra_neuman2011} when leakage effects are minor, but our solution predicts less drawdown when leakage effects are significant. It is seen in Figure~\ref{fig2}b that solution of \citet{mishra_neuman2011} overestimates drawdown near the aquitard at intermediate time because it does not include aquitard leakage.  Near the water table (Figure~\ref{fig2}a) the effects of aquitard leakage are minimal and our proposed solution approaches \citet{mishra_neuman2011} at all times.

\begin{figure}
\begin{center}
\includegraphics[width = 8 cm]{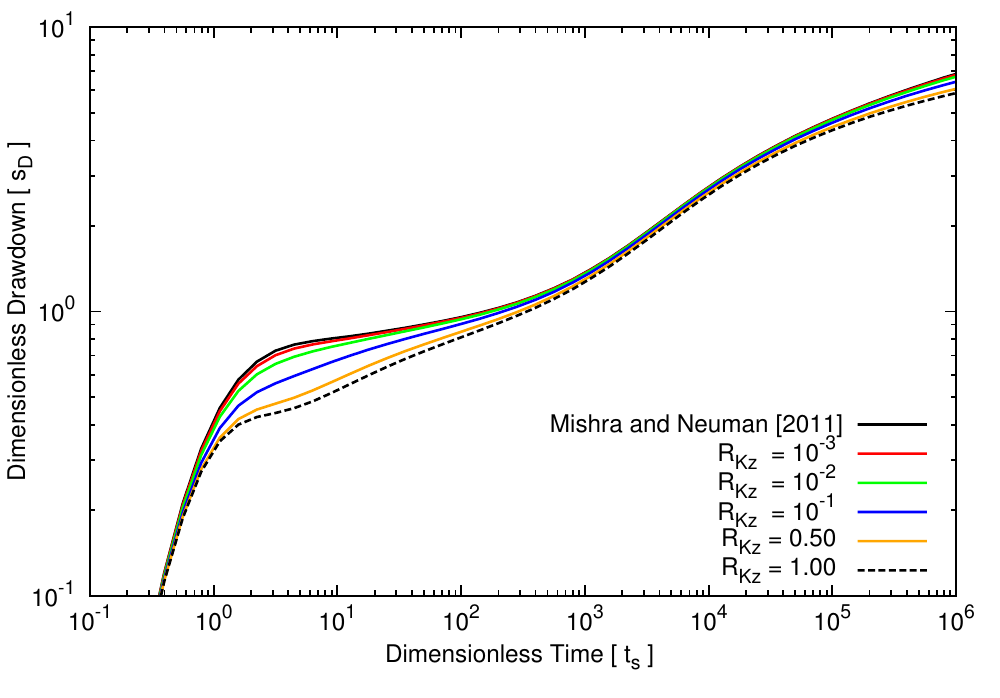}
\includegraphics[width = 8 cm]{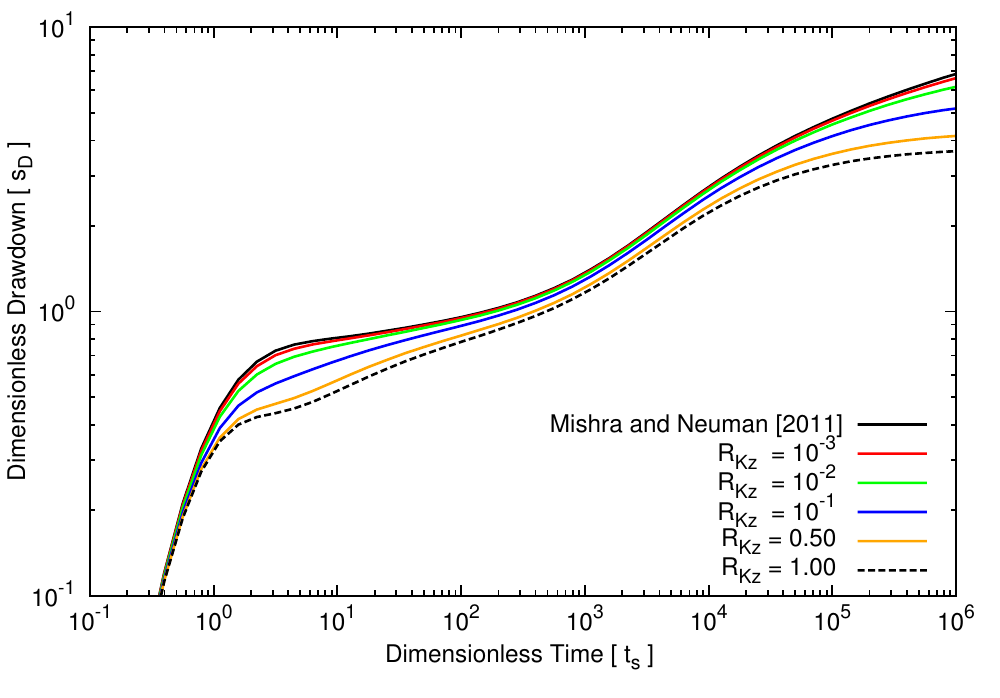}
\caption[Time-drawdown at various depths]{Dimensionless
  leaky-unconfined aquifer drawdown versus dimensionless time at
  $r_D=0.5$  and $z_D=0.25$ for $K_D=1$, $S_Sb/S_y=10^{-3}$,
  $a_{kD}=a_{cD}=10$, $\psi_{aD}=\psi_{kD}$, $d_D=0$, $l_D=0.6$,
  $C_{wD}=10^2$, $R_{S_s}=10^{-2}$, $R_b \to \infty$ when $R_{K_z}$
  varies and (a) $R_{K_R}=10^{-6}$ (b) $R_{K_r}=1$. Solution of
  \citet{mishra_neuman2011} is also shown. \label{fig3}}
\end{center}
\end{figure}

Figures~\ref{fig3}a and \ref{fig3}b show dimensionless time-drawdown variations at dimensionless radial distance $r_D=0.5$  and dimensionless unconfined aquifer saturated zone elevation $z_D=0.25$ with different values of $R_{K_z}=K_{z1}/K_z$  when the radial aquitard hydraulic conductivity is small ($R_{K_r}=K_{r1}/K_r=10^{-6}$) and large ($R_{k_r}=1.0$). When the radial hydraulic conductivity in aquitard is negligible ($R_{k_r}=10^{-6}$), aquitard flow is predominately vertical; larger values of vertical aquitard hydraulic conductivity cause decreases in intermediate time drawdown (Figure~\ref{fig3}a). It is seen from Figure~\ref{fig3}b that when aquitard horizontal hydraulic conductivity is large ($R_{k_r}=1$) the amount drawdown is reduced from further increases in aquitard vertical hydraulic conductivity also extend to the later time.

\begin{figure}
\begin{center}
\includegraphics[width = 8 cm]{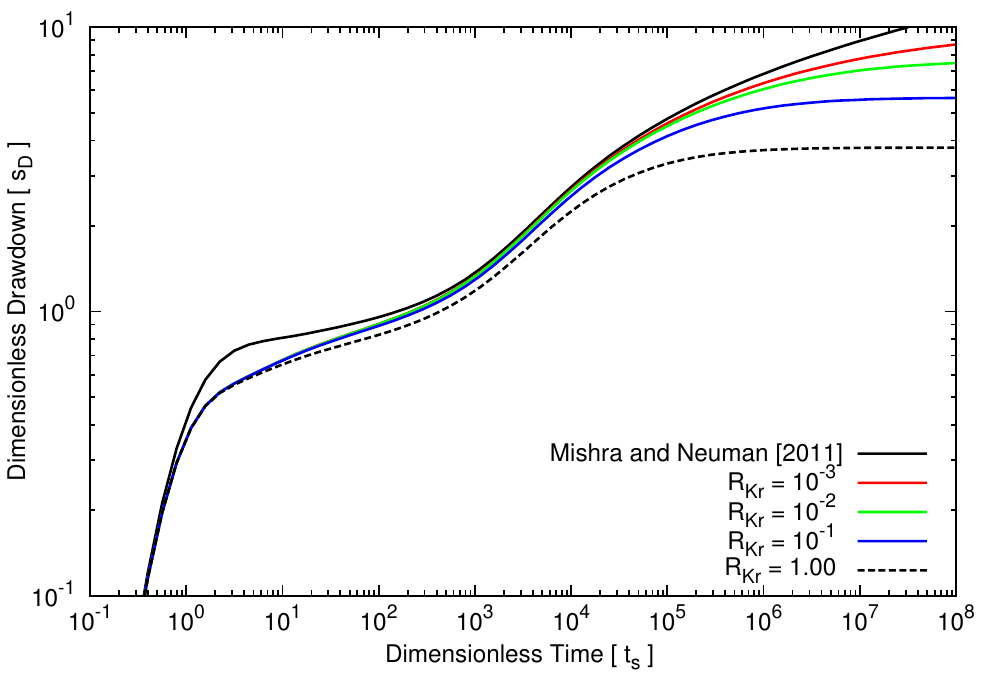}
\caption[Time-drawdown at various depths]{Dimensionless
  leaky-unconfined aquifer drawdown versus dimensionless time at
  $r_D=0.5$  and $z_D=0.25$ for $K_D=1$, $S_Sb/S_y=10^{-3}$,
  $a_{kD}=a_{cD}=10$, $\psi_{aD}=\psi_{kD}$, $d_D=0$, $l_D=0.6$,
  $C_{wD}=10^2$, $R_{S_s}=10^{-2}$, $R_{K_z}=0.1$ and $R_b \to \infty$
  when $R_{K_r}$ varies. Solution of \citet{mishra_neuman2011} is
  also shown. \label{fig4}}
\end{center}
\end{figure}

Figure~\ref{fig4} depicts the effect of $R_{K_r}$  on the dimensionless time-drawdown at dimensionless radial distance $r_D=0.5$ and dimensionless unconfined aquifer saturated zone elevation $z_D=0.25$   when $R_{K_z}=0.1$. Radial flow in the aquitard results in less drawdown at late time than that predicted by \citet{mishra_neuman2011}, who do not account for aquitard leakage.

\begin{figure}
\begin{center}
\includegraphics[width = 8 cm]{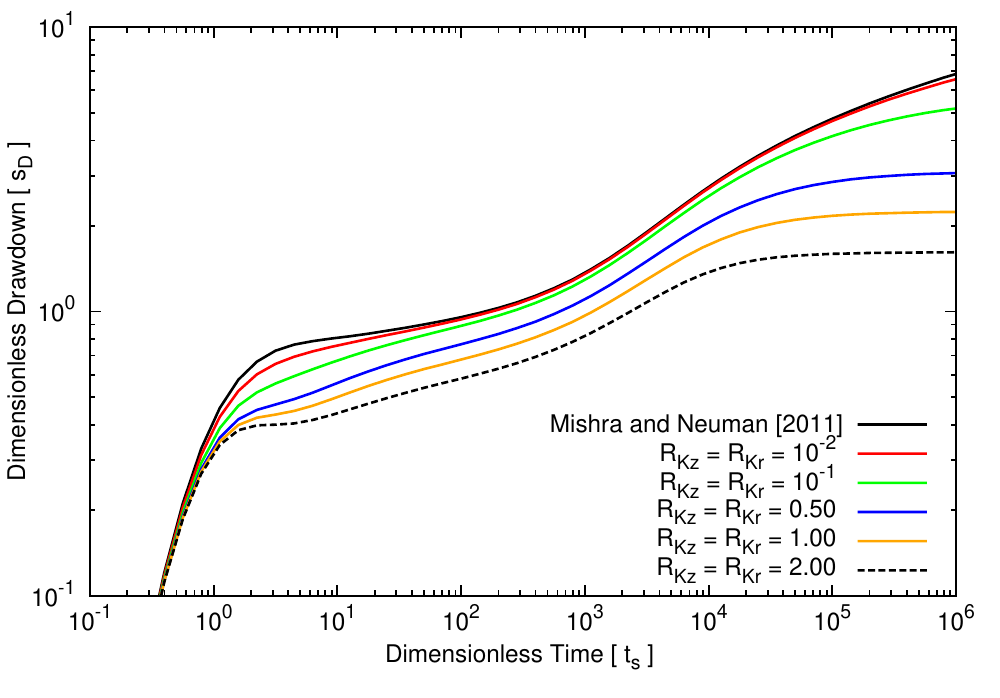}
\caption[Time-drawdown at various depths]{Dimensionless
  leaky-unconfined aquifer drawdown versus dimensionless time at
  $r_D=0.5$  and $z_D=0.25$ for $K_D=1$, $S_Sb/S_y=10^{-3}$,
  $a_{kD}=a_{cD}=10$, $\psi_{aD}=\psi_{kD}$, $d_D=0$, $l_D=0.6$,
  $C_{wD}=10^2$, $R_b \to \infty$  when $R_{K_z}=R_{K_r}$ varies and
  $R_{S_s}=1$. Solution of \citet{mishra_neuman2011} is also shown. \label{fig5}}
\end{center}
\end{figure}

Figure~\ref{fig5} presents the effect of hydraulic conductivity of an
isotropic aquitard on dimensionless time-drawdown at dimensionless
radial distance $r_D=0.5$  and dimensionless unconfined aquifer
saturated zone elevation $z_D=0.25$. When aquitard hydraulic
conductivity is at least two orders of magnitude smaller than the
unconfined aquifer, the effects of leakage on the aquifer drawdown are
negligible. This is in agreement with findings of \citet{neuman1969a}
for confined aquifers.  They found errors $<5\%$ attributable to the
vertical aquitard flow assumption, when the hydraulic conductivity
contrast between the aquifer and aquitard was greater than a factor of
$100$. Figure~\ref{fig5} also presents a case with the aquitard hydraulic conductivity is larger than the aquifer. Because the proposed model accounts for general three-dimensional flow in underlying zone, we can consider the case where the lower layer is more permeable than the aquifer ($R_{k_r}=2$).

\begin{figure}
\begin{center}
\includegraphics[width = 8 cm]{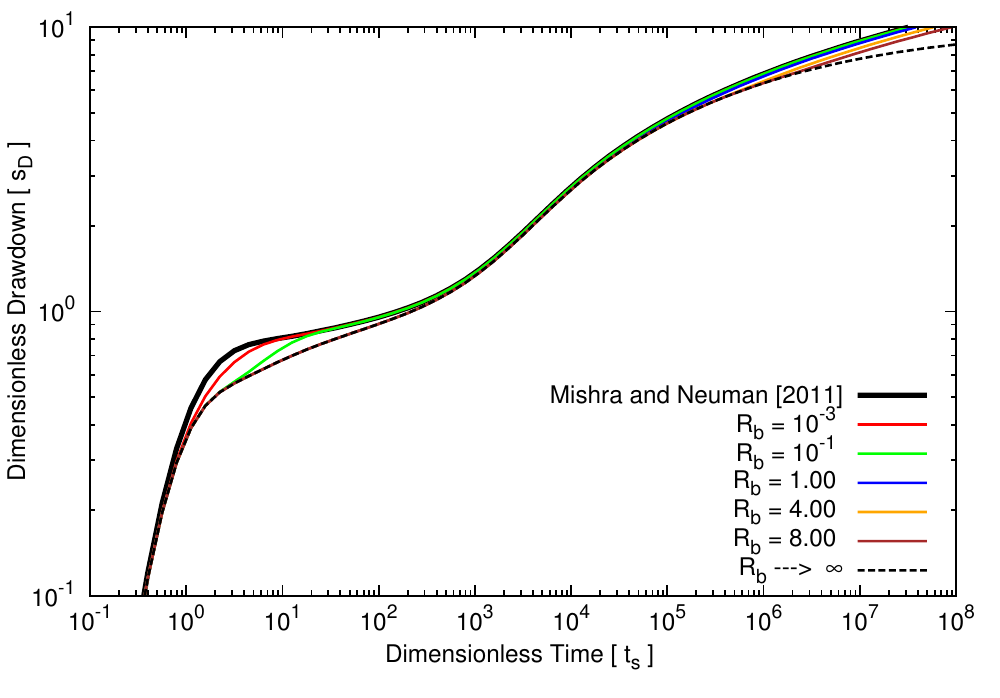}
\caption[Time-drawdown at various depths]{Dimensionless
  leaky-unconfined aquifer drawdown versus dimensionless time at
  $r_D=0.5$  and $z_D=0.25$ for $K_D=1$, $S_Sb/S_y=10^{-3}$,
  $a_{kD}=a_{cD}=10$, $\psi_{aD}=\psi_{kD}$, $d_D=0$, $l_D=0.6$,
  $C_{wD}=10^2$, $R_{S_s}=10^2$, $R_{K_z}=R_{K_r}=10^{-2}$  when
  $R_b=b_1/b$ varies. Solution of \citet{mishra_neuman2011} is also shown. \label{fig6}}
\end{center}
\end{figure}

Figure~\ref{fig6} shows how the dimensionless unconfined aquifer
time-drawdown is affected by aquitard thickness. When the aquitard
thickness is less than the initial unconfined aquifer saturated
thickness ($R_b\le 1$) aquitard leakage only affects the time-drawdown
curve at intermediate time. Figure~\ref{fig6} shows that further increases in aquitard thickness beyond eight times the initial unconfined aquifer saturated zone thickness have negligible effect on the time-drawdown curve.

\subsection{Dimensionless aquitard time-drawdown}
\begin{figure}
\begin{center}
\includegraphics[width = 8 cm]{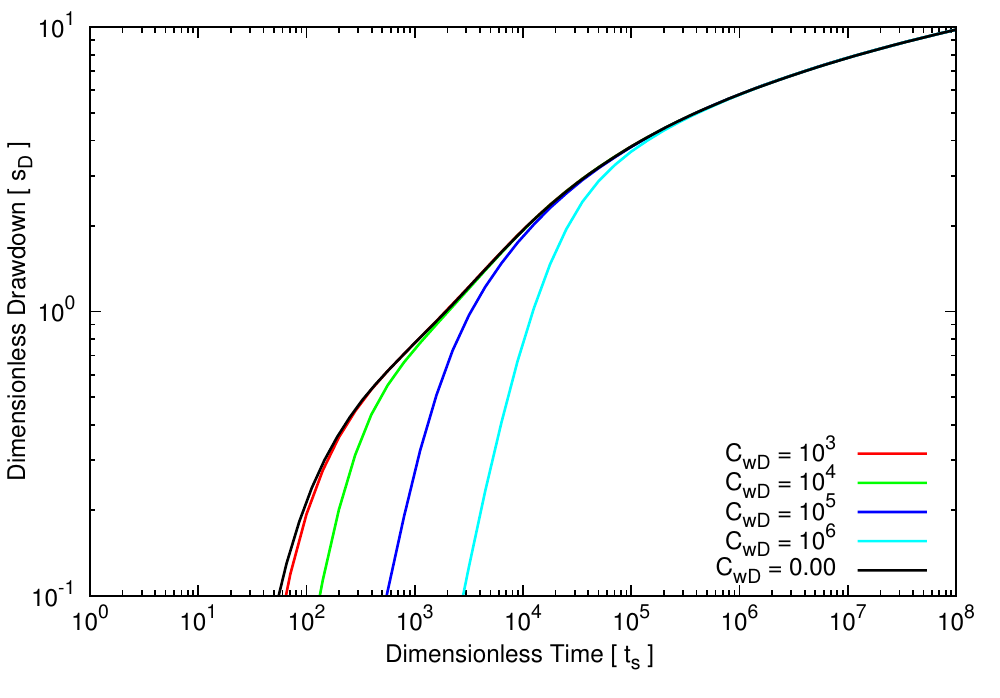}
\caption[Time-drawdown at various depths]{Dimensionless aquitard drawdown versus dimensionless time at $r_D=0.2$ and $z_D=-0.25$ for $K_D=1$, $S_Sb/S_y=10^{-3}$, $a_{kD}=a_{cD}=10$, $\psi_{aD}=\psi_{kD}$, $d_D=0$, $l_D=0.6$, $R_{S_s}=10^2$, $R_{K_z}=R_{K_r}=10^{-2}$, $R_b \to \infty$ when $C_{wD}$, the dimensionless wellbore storage varies. \label{fig7}}
\end{center}
\end{figure}

Figure~\ref{fig7} depicts dimensionless aquitard drawdown $s_D(r_D,z_d,t_s)=(4\pi K_rb/Q)s_1(r_D,z_D,t_s)$   variations with dimensionless time at dimensionless radial distance $r_D=0.2$  and dimensionless aquitard elevation  $z_D=-0.25$ for different values of $C_{wD}$ . As with solution of \citet{mishra_neuman2011} for non-leaky systems, aquitard drawdown is impacted by pumping-well wellbore storage capacity. Larger wellbore storage factors result in increased capacity of the wellbore to store water, resulting in a delay in the aquitard time-drawdown, as indicated in Figure 7.

\begin{figure}
\begin{center}
\includegraphics[width = 8 cm]{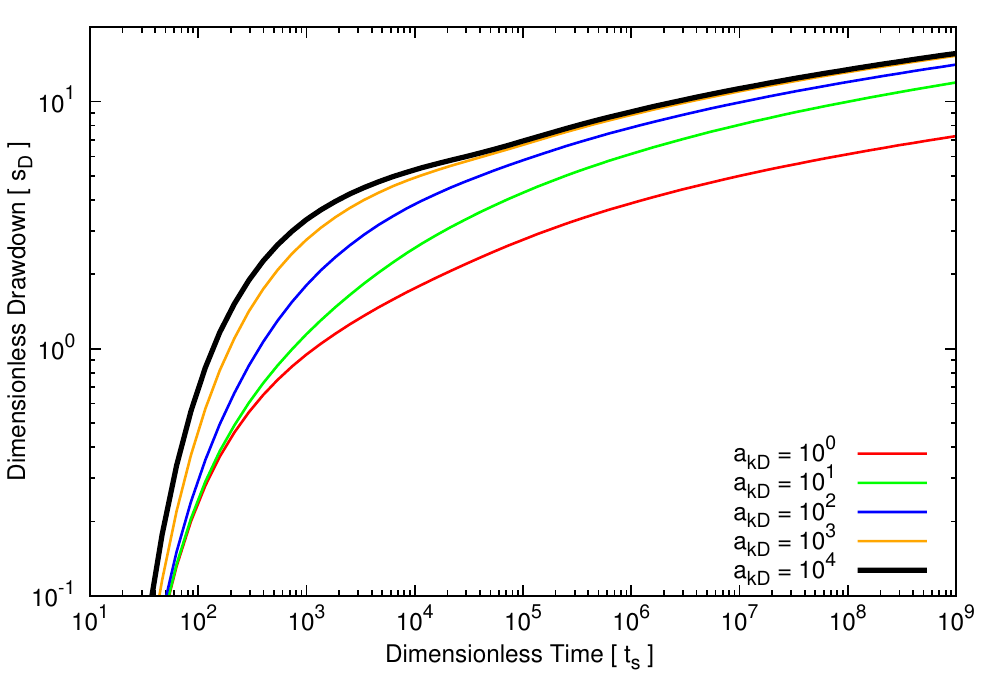}
\caption[Time-drawdown at various depths]{Dimensionless aquitard drawdown versus dimensionless time at $r_D=0.2$ and $z_D=-0.25$ for $K_D=1$, $S_Sb/S_y=10^{-3}$, $\psi_{aD}=\psi_{kD}$, $d_D=0$, $l_D=0.6$,  $C_{wD}=10^2$, $R_{S_s}=10^2$, $R_{K_z}=R_{K_r}=10^{-2}$, $R_b \to \infty$ when, $a_{cD}=1$ and $a_{kD}$ varies. \label{fig8}}
\end{center}
\end{figure}

Figure~\ref{fig8} depicts the effect that changes in $a_{kD}$, the dimensionless relative hydraulic conductivity exponent, have on dimensionless time-drawdown at dimensionless radial distance $r_D=0.2$ and dimensionless aquitard elevation $z_D=-0.25$. For larger values of $s_{kD}$, the unsaturated zone hydraulic conductivity decreases more rapidly as pressure becomes more negative, relative to the threshold pressure $\psi_k$ . A diminishing rate of water then drains from the vadose zone into the aquifer; this drainage contributes to reduced aquitard drawdown. For very large $a_{kD}$, unsaturated hydraulic conductivity quickly decreases once pressure head is below $\psi_k$ , which leads to an much less permeable unsaturated zone.

\begin{figure}
\begin{center}
\includegraphics[width = 8 cm]{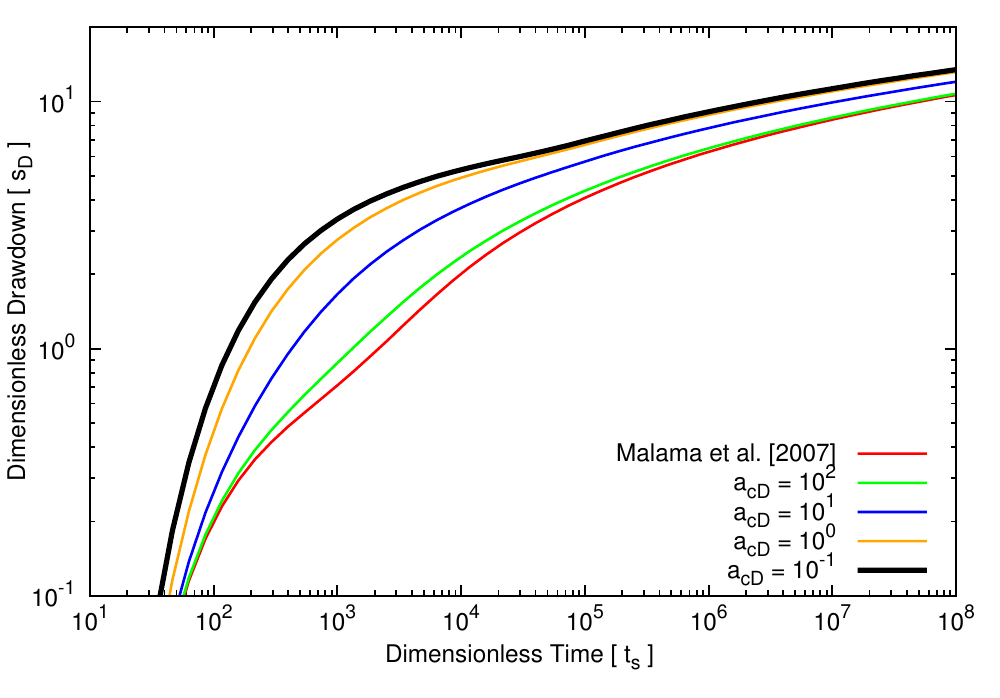}
\caption[Time-drawdown at various depths]{Dimensionless aquitard drawdown versus dimensionless time at $r_D=0.2$ and $z_D=-0.25$ for $K_D=1$, $S_Sb/S_y=10^{-3}$, $\psi_{aD}=\psi_{kD}$, $d_D=0$, $l_D=0.6$,  $C_{wD}=10^2$, $R_{S_s}=10^2$, $R_{K_z}=R_{K_r}=10^{-2}$, $R_b \to \infty$ when, $a_{kD}=10^3$ and $a_{cD}$ varies. \label{fig9}}
\end{center}
\end{figure}

Figure~\ref{fig9} shows the effects that changes in $a_{cD}$, the dimensionless effective saturation exponent, have on dimensionless time-drawdown at dimensionless radial distance $r_D=0.2$  and dimensionless aquitard elevation $z_D=-0.25$ . When $a_{cD}$ and $a_{kD}$ are both large, pressure head and hydraulic conductivity in the vadose zone quickly reduce as pressure reaches the thresholds $\psi_k$  and $\psi_a$. The vadose zone can no longer store water, and the water table essentially becomes a moving boundary, which leads to the limiting-case behavior of instantaneous drainage due to \citet{neuman1972}. Consequently, for large values of exponents (Figure~\ref{fig9}, red curve) the proposed solution reduces to that of \citet{malama2007}, which relies on the assumption of instantaneous drainage of \citet{neuman1972}. As $a_{cD}$  decreases,the vadose zone has increased capacity to store water, which diminishes the water table response  and aquifer drawdown increases, compared to that predicted by \citet{malama2007}.

\section{Conclusions}
Our work leads to the following major conclusions:
\begin{enumerate}
\item A new analytical solution was developed for axially symmetric saturated-unsaturated three dimensional radial flow to a well with wellbore storage that partially penetrates the saturated zone of a compressible vertically anisotropic leaky-unconfined aquifer. The solution accounts for both radial and vertical flow in the unsaturated zone and the underlying aquitard.
\item Because the solution considers three-dimensional radial flow in the aquitard, any properties may be assigned to the aquitard, allowing the solution to also be used to simulate leakage from underlying non-aquitard layers (e.g., an unscreened aquifer region with different hydraulic properties).
\item Aquitard leakage can lead to significant departures from solutions that do not account for leakage, e.g., \citet{mishra_neuman2011}. However, the effect of leakage on unconfined aquifer drawdown diminishes at points farther away from the aquifer-aquitard boundary.
\item Unsaturated zone effects are often more important than leakage effects when the observation location is close to the water table.
\item For large diameter pumping wells, at early time water is withdrawn entirely from the wellbore storage. Solution that do not account for wellbore storage predict a much larger early rise in drawdown.
\item Aquitard drawdown is also affected by the pumping-well wellbore storage capacity. As in the unconfined aquifer, larger wellbore storage capacity leads to larger impacts on the observed aquitard drawdown.
\item The unsaturated zone properties not only affect the unconfined aquifer time-drawdown behavior but they also impact the observed aquitard response.
\end{enumerate}

\section*{Acknowledgments}
This research was partially funded by the Environmental Programs Directorate of the Los Alamos National Laboratory. Los Alamos National Laboratory is a multi-program laboratory managed and operated by Los Alamos National Security (LANS) Inc. for the U.S. Department of Energy’s National Nuclear Security Administration under contract  DE-AC52-06NA25396.  Sandia National Laboratories is a multi-program laboratory managed and operated by Sandia Corporation, a wholly owned subsidiary of Lockheed Martin Corporation, for the U.S. Department of Energy’s National Nuclear Security Administration under contract DE-AC04-94AL85000.

\newpage

\begin{table}
\caption{\bf{Fundamental Properties Table}}
\footnotesize
\begin{tabular}{|l|l|l|}
\hline
$a$     &   Hankel transform parameter  &   $L^{-1}$    \\
$a_c$   &   exponent in moisture retention curve or sorptive number     &   $L^{-1}$    \\
$a_k$   &   exponent in Gardner relative hydraulic conductivity model   &   $L^{-1}$    \\
$b$     &   saturated thickness of unconfined aquifer before pumping begins     &   $L$     \\
$b_1$   &   thickness of aquitard   &   $L$     \\
$C_w$   &   wellbore storage coefficient    &   $L^2$   \\
$d$     &   distance from top of screened interval to top of aquifer    &   $L$     \\
$h$     &   hydraulic head (sum of pressure and elevation heads)    &   $L$     \\
$K_r$   &   aquifer radial hydraulic conductivity   &   $LT^{-1}$   \\
$K_{r1}$   &   aquitard radial hydraulic conductivity  &   $LT^{-1}$   \\
$K_z$   &   aquifer vertical hydraulic conductivity     &   $LT^{-1}$   \\
$K_{z1}$   &   aquitard vertical hydraulic conductivity    &   $LT^{-1}$   \\
$l$     &   distance from bottom of screened interval to top of aquifer     &   $L$     \\
$L$     &   thickness of vadose zone before pumping begins  &   $L$     \\
$n$     &   finite cosine transform parameter   &   $-$     \\
$p$     &   Laplace transform parameter     &   $T^{-1}$    \\
$Q$     &   volumetric pumping rate     &   $L^3T^{-1}$     \\
$r$     &   radial distance from the center of pumping well     &   $L$     \\
$r_w$   &   diameter of pumping well    &   $L$     \\
$s$     &   drawdown in aquifer; change in hydraulic head since pumping began   &   $L$     \\
$s_1$   &   drawdown in aquitard; change in hydraulic head since pumping began  &   $L$     \\
$S_e$   &   effective saturation    &   $-$     \\
$S_s$   &   aquifer specific storage    &   $L^{-1}$    \\
$S_{s1}$   &   aquitard specific storage   &   $L^{-1}$    \\
$S_y$   &   aquifer drainable porosity or specific yield    &   $-$     \\
$t$     &   time since pumping began    &   $T$     \\
$z$     &   vertical distance from the bottom of the aquifer, positive up   &   $L$     \\
$z_i$    &   elevation to top ($i=1$) and bottom ($i=2$) of monitoring interval   &   $L$     \\
$\theta_0$  &   initial volumetric water content    &   $-$     \\
$\theta_r$  &   residual volumetric water content   &   $-$     \\
$\theta_s$  &   saturated volumetric water content  &   $-$     \\
$\sigma$    &   drawdown in unsaturated zone; change in hydraulic head since pumping began  &   $L$     \\
$\psi$  &   pressure head (less than zero when unsaturated)     &   $L$     \\
$\psi_a$    &   air-entry pressure  &   $L$     \\
$\psi_k$    &   pressure for saturated hydraulic conductivity   &   $L$     \\
\hline
\end{tabular}
\end{table}

\begin{table}
\caption{\bf{Derived quantities table}}
 \footnotesize
\begin{tabular}{|l|l|l|}
\hline
$K_D$ &     $K_z/K_r$   &   Anisotropy ratio    \\
$r_D$ &     $r/b$   &   dimensionless radial coordinate     \\
$z_D$ &     $z/b$   &   dimensionless vertical coordinate   \\
$d_D$ &     $d/b$   &   dimensionless distance to top of screen interval    \\
$l_D$ &     $l/b$   &   dimensionless distance to bottom of screen interval     \\
$p_D$ &     $pt$    &   dimensionless Laplace parameter     \\
$r_{w_D}$   &    $r_w/r$     &   dimensionless well radius   \\
$R_{K_D}$   &    $K_{D1}/K_D$   &   ratio of aquitard and aquifer
anistropies   \\
$R_{K_r}$  &    $K_{r1}/K_r    $&  ratio of aquitard and aquifer
horizontal hydraulic conductivities     \\
$R_{K_z}$  &    $K_{z1}/K_z    $&  ratio of aquitard and aquifer vertical hydraulic conductivities     \\
$R_{\alpha_s}$  &   $\alpha_{s1}/\alpha_s$     &   ratio of aquitard and aquifer saturated hydraulic diffusivities     \\
$R_b$ &     $b_1/b$     &   ratio of aquitard and aquifer thicknesses   \\
$\alpha_s$  &   $K_r/S_s$   &   aquifer hydraulic diffusivity   \\
$\alpha_{s1}$  &   $K_{r1}/S_{s1}$   &   aquitard hydraulic diffusivity  \\
$z_{D_i}$   &   $z_i/b$     &  dimensionless elevation to top ($i=1$) and bottom ($i=2$) of monitoring interval   \\
$a_{kD}$    &   $a_kb$  &   dimensionless Gardner hydraulic conductivity model exponent     \\
$a_{cD}$    &   $a_cb$  &   dimensionless moisture retention model exponent     \\
$\psi_{aD}$     &   $\psi_a/b$  &   dimensionless air-entry pressure    \\
$\psi_{kD}$     &   $\psi_k/b$  &   dimensionless pressure for saturated hydraulic conductivity     \\
$C_{wD}$    &   $C_w/(\pi S_sbr_w^2)$   &   dimensionless wellbore storage coefficient  \\
$t_s$ &     $\alpha st/r^2$     &   dimensionless time  \\
\hline
\end{tabular}
\end{table}

\newpage

\appendix
\newpage
\section{Decomposition of saturated zone solution}
In a manner analogous to \citet{mishra_neuman2010} we decompose $s$ into two parts
\begin{equation}
s={{s}_{C}}+{{s}_{U}}
\end{equation}
where ${{s}_{C}}$ is solution for a partially penetrating well in a confined aquifer, satisfying
\begin{eqnarray}
\frac{1}{r}\frac{\partial }{\partial r}\left( r\frac{\partial {{s}_{C}}}{\partial r} \right)+{{K}_{D}}\frac{{{\partial }^{2}}{{s}_{C}}}{\partial {{z}^{2}}}=\frac{1}{{{\alpha }_{s}}}\frac{\partial {{s}_{C}}}{\partial t} & r\ge {{r}_{w}} & 0\le z<b
\end{eqnarray}
\begin{eqnarray}
{{s}_{C}}(r,z,0)=0 && r\ge {{r}_{w}}
\end{eqnarray}
\begin{equation}
{{s}_{C}}\left( \infty ,z,t \right)=0
\end{equation}
\begin{eqnarray}\label{confined_bound}
\frac{\partial s_C}{\partial z}\bigg|_{z=(0,b)}=0 && r\ge r_w
\end{eqnarray}
\begin{eqnarray}\nonumber
{{\left( \frac{\partial {{s}_{C}}}{\partial r} \right)}_{r={{r}_{w}}}}=0
&& 0\le z\le b-l \quad b-d\le z\le b \\
\end{eqnarray}
\begin{eqnarray}\nonumber
2\pi {{K}_{r}}\left( l-d \right){{\left. \frac{\partial {{s}_{C}}}{\partial r} \right|}_{r={{r}_{w}}}}-{{C}_{W}}{{\left. \frac{\partial {{s}_{C}}}{\partial t} \right|}_{r={{r}_{w}}}}=-Q \\ \nonumber
&& b-l\le z\le b-d \\
\end{eqnarray}
and ${{s}_{U}}$ is a solution that takes into account aquitard and saturated-unsaturated unconfined conditions, but has no pumping source term, satisfying
\begin{eqnarray}\label{su_begin}
\frac{1}{r}\frac{\partial }{\partial r}\left( r\frac{\partial {{s}_{U}}}{\partial r} \right)+{{K}_{D}}\frac{{{\partial }^{2}}{{s}_{U}}}{\partial {{z}^{2}}}=\frac{1}{{{\alpha }_{s}}}\frac{\partial {{s}_{U}}}{\partial t} \\ \nonumber
r\ge 0 && 0\le z<b
\end{eqnarray}
\begin{equation}
{{s}_{U}}(r,z,0)=0 \quad r\ge 0
\end{equation}
\begin{equation}
{{s}_{U}}\left( \infty ,z,t \right)=0
\end{equation}
\begin{eqnarray}
\frac{\partial s_U}{\partial z}-\frac{K_{z1}}{K_z}\frac{\partial s_1}{\partial z}=0 & r\ge 0 & z=0
\end{eqnarray}
\begin{eqnarray}\label{su_end}
\frac{\partial s_U}{\partial z}\bigg|_{r=0}=0 && 0\le z\le b
\end{eqnarray}
subject to interface conditions at water table,
\begin{eqnarray}\label{interface1}
{{s}_{C}}+{{s}_{U}}-\sigma =0 \quad r\ge {{r}_{w}} && z=b
\end{eqnarray}
\begin{eqnarray}\label{interface2}
\frac{\partial {{s}_{C}}}{\partial z}+\frac{\partial {{s}_{U}}}{\partial z}-\frac{\partial \sigma }{\partial z}=0 \quad r \ge {{r}_{w}} && z=b,
\end{eqnarray}
where the first term is zero by definition of $s_C$.

\section{Laplace-space solution for saturated zone}
Equations \eqref{su_begin}--\eqref{interface2} are solved by sequential application of the Hankel transform
\begin{equation}
f(a)=\int\limits_{0}^{\infty }{r{\mathrm{J}_0}(ar)f(r) \; \mathrm{d}r}
\end{equation}
and Laplace transform
\begin{equation}
f(p)=\int\limits_{0}^{\infty }{f(t){{e}^{-pt}}\; \mathrm{d}t}
\end{equation}
with Hankel parameter $a$ and Laplace parameter $p$, $\mathrm{J}_0$ being zero-order Bessel function of the first kind.

The Laplace-Hankel transform of confined aquifer solution \citep{mishra_neuman2011} is
\begin{eqnarray}\nonumber
\bar{\bar{s}}_C(a,z_D,p_D)&=& C_0 \left\{ \frac{r_w}{a}\mathrm{J}_1(ar_w)\mathrm{K}_0(r_w\tau_0) \right. \\ \nonumber
&& \left. +\frac{\tau_0r_w\mathrm{J}_0(ar)\mathrm{K}_1(r\tau_0)-ar_w\mathrm{J}_1(ar)\mathrm{K}_0(r\tau_0)}{a^2+\tau_0^2}\right\} \\ \nonumber
&& +\sum_{n=1}^\infty C_n\left\{\frac{r_w}{a}\mathrm{J}_1(ar_w)\mathrm{K}_0(r_w\tau_0) \right. \\
&&\left. +\frac{\tau_nr_w\mathrm{J}_0(ar)\mathrm{K}_1(r\tau_n)-ar_w\mathrm{J}_1(ar)\mathrm{K}_0(r\tau_n)}{a^2+\tau_n^2}\right\} \\ \nonumber
&& \times\cos\left[n\pi (1-z_D)\right]
\end{eqnarray}
where $\tau_0=\sqrt{pS_s/K_r}$ and $\tau_n=\sqrt{pS_s/K_r+K_Dn^2\pi^2/b^2}$.

The Laplace transform of \eqref{su_begin}--\eqref{interface2} is
\begin{eqnarray}\label{su_lap_begin}
\frac{1}{r}\frac{\partial }{\partial r}\left( r\frac{\partial {{{\bar{s}}}_{U}}}{\partial r} \right)+{{K}_{D}}\frac{{{\partial }^{2}}{{{\bar{s}}}_{U}}}{\partial {{z}^{2}}}=\frac{p}{{{\alpha }_{s}}}{{\bar{s}}_{U}} && 0\le z<b
\end{eqnarray}
\begin{equation}
{{\bar{s}}_{U}}\left( \infty ,z,p \right)=0
\end{equation}
\begin{equation}
\frac{\partial \bar{s}_U}{\partial z}\bigg|_{z=0}=\frac{K_{z1}}{K_z}\frac{\partial \bar{s}_1}{\partial z}
\end{equation}
\begin{eqnarray}
\left(r\frac{\partial \bar{s}_U}{\partial r}\right)_{r=0}=0 & 0 \le z \le b
\end{eqnarray}
\begin{eqnarray}
\bar{s}_C+\bar{s}_U-\bar{\sigma}=0 && z=b
\end{eqnarray}
\begin{eqnarray}\label{su_lap_end}
\frac{\partial {{{\bar{s}}}_{C}}}{\partial z}+\frac{\partial {{{\bar{s}}}_{U}}}{\partial z}-\frac{\partial \bar{\sigma }}{\partial \text{z}}=0 &&  z=b
\end{eqnarray}
where the first term is zero by definition of $\bar{s}_C$.

Taking the Hankel transform of \eqref{su_lap_begin}--\eqref{su_lap_end} yields
\begin{eqnarray}\label{su_hankel1}
-{{a}^{2}}{{\bar{\bar{s}}}_{U}}+{{K}_{D}}\frac{{{\partial }^{2}}{{{\bar{\bar{s}}}}_{U}}}{\partial {{z}^{2}}}=\frac{p}{{{\alpha }_{s}}}{{\bar{\bar{s}}}_{U}} &&     0\le z<b
\end{eqnarray}
\begin{eqnarray}\label{su_hankel2}
{{\bar{\bar{s}}}_{H}}+{{\bar{\bar{s}}}_{U}}-\bar{\bar{\sigma }}=0 && z=b
\end{eqnarray}
\begin{eqnarray}\label{su_hankel3}
\frac{\partial {{{\bar{\bar{s}}}}_{H}}}{\partial z}+\frac{\partial {{{\bar{\bar{s}}}}_{U}}}{\partial z}-\frac{\partial \bar{\bar{\sigma }}}{\partial \text{z}}=0 && z=b
\end{eqnarray}
\begin{eqnarray}\label{su_hankel4}
\frac{\partial {{{\bar{\bar{s}}}}_{C}}}{\partial z}+\frac{\partial {{{\bar{\bar{s}}}}_{U}}}{\partial z}-\frac{{{K}_{z1}}}{{{K}_{z}}}\frac{\partial {{{\bar{\bar{s}}}}_{1}}}{\partial \text{z}}=0 && z=0
\end{eqnarray}
The general solution of \eqref{su_hankel1} subject to \eqref{su_hankel2} is
\begin{equation}\label{su_sol_hank}
{{\bar{\bar{s}}}_{U}}={{\rho }_{1}}{{e}^{\eta z}}+{{\rho }_{2}}{{e}^{-\eta z}}
\end{equation}
where $\rho_1$ and $\rho_2$ are coefficients to be determined from boundary conditions.

Considering that $ \partial \bar{\bar{s}}_H/\partial z =0$ at  $z=0$ and $z=b$ by virtue of \eqref{confined_bound} and that
\begin{equation}
{{\left. \frac{\partial \bar{\bar{\sigma }}}{\partial z} \right|}_{z=b}}=q{{\left( {{{\bar{\bar{s}}}}_{C}}+{{{\bar{\bar{s}}}}_{U}} \right)}_{z=b}}
\end{equation}
which, together with $q$, are derived in (D15) of \citet{mishra_neuman2011} and
\begin{equation}
{{\left. \frac{\partial {{{\bar{\bar{s}}}}_{U}}}{\partial z} \right|}_{z=0}}=q_1{{\left( {{{\bar{\bar{s}}}}_{C}}+{{{\bar{\bar{s}}}}_{U}} \right)}_{z=0}}
\end{equation}
which, together with $q_1$ , are derived in \eqref{aquitard_flux} we obtain from \eqref{su_hankel2}--\eqref{su_hankel4}
\begin{equation}
\rho_1=\frac{q_1(\eta+q)e^{-\eta b}\bar{\bar{s}}_C(z=0)-q(\eta+q_1)\bar{\bar{s}}(z=b)}{\Delta}
\end{equation}
\begin{equation}
\rho_2=\frac{q_1(\eta-q)e^{-\eta b}\bar{\bar{s}}_C(z=0)-q(\eta-q_1)\bar{\bar{s}}(z=b)}{\Delta}
\end{equation}
where $\Delta =(\eta-q_1)(\eta+q)e^{-\eta b}-(\eta-q)(\eta+q_1)e^{\eta b}$.

The inverse Hankel transform of \eqref{su_sol_hank} is
\begin{equation}\label{su_gen_lap}
\bar{s}_U=\int_0^\infty \left(\rho_1e^{\eta z}+\rho_2e^{-\eta
    z}\right)a\mathrm{J}_0(ar) \;\mathrm{d}a.
\end{equation}
Defining a new variable $y=ar/K_D^{1/2}r_D$ transforms \eqref{su_gen_lap} into the result presented in \eqref{su_final}.
It is noted that when $q_1=0$  the aquiard is replaced by an impermeable boundary, and $\rho_1=\rho_2=\frac{2\bar{\bar {s}}_C(z=b)}{\cosh(\eta b)-\frac{\eta}{q}\sinh(\eta b)}$ . These simplifications reduce \eqref{su_gen_lap} to equation (3) of \citet{mishra_neuman2011}.

\section{Aquitard Solution}
Laplace--Hankel transform of governing flow equations for aquitard  are
\begin{equation}\label{aquitard_double_trans}
-{{a}^{2}}{{\bar{\bar{s}}}_{1}}+{{K}_{D1}}\frac{{{\partial }^{2}}{{{\bar{\bar{s}}}}_{1}}}{\partial {{z}^{2}}}=p\frac{{{S}_{s1}}}{{{K}_{r1}}}{{\bar{\bar{s}}}_{1}} \quad 0\le z<-{{b}_{1}}
\end{equation}
By virtue of no flow boundary at the bottom of the system, ${{\left. \frac{\partial {{{\bar{\bar{s}}}}_{1}}}{\partial z} \right|}_{z=-{{b}_{1}}}}=0$, the general solution to \eqref{aquitard_double_trans} is
\begin{equation}
{{\bar{\bar{s}}}_{1}}={{\rho }_{1}}\cosh \left[ {{\eta }_{1}}\left( z+{{b}_{1}} \right) \right]
\end{equation}
where $\eta_1^2=\frac{a^2}{K_{D1}}+\frac{pS_{s1}}{K_{r1}K_{D1}}$.
The boundary condition
\begin{equation}
{{ {{{\bar{\bar{s}}}}_{1}} }(z=0)}={{ {\bar{\bar{s}}} }(z=0)}={{\left( {{{\bar{\bar{s}}}}_{C}}+{{{\bar{\bar{s}}}}_{U}} \right)}_{z=0}}
\end{equation}
gives
\begin{equation}\label{aquitard_hankel_gen}
s_1=\frac{(\bar{\bar{s}}_C+\bar{\bar{s}}_U)_{z=0}}{\cosh(\eta b_1)}\cosh\left[\eta_1 (z+b_1)\right].
\end{equation}
Using \eqref{su_hankel4} transforms \eqref{aquitard_hankel_gen} into the solution presented in \eqref{aquitard_drawdown}.

The derivative of \eqref{aquitard_hankel_gen} is
\begin{eqnarray}
\frac{d\bar{\bar{s}}_1}{dz}=\eta_1 \tanh(\eta_1b_1)(\bar{\bar{s}}_C+\bar{\bar{s}}_U) && z=0.
\end{eqnarray}
The flux boundary condition at the aquifer-aquitard interface
\begin{equation}
{{\left. \frac{\partial {{{\bar{\bar{s}}}}_{1}}}{\partial z} \right|}_{z=0}}={{\left. \frac{{{K}_{z}}}{{{K}_{z1}}}\frac{\partial \bar{\bar{s}}}{\partial z} \right|}_{z=0}}
\end{equation}
Combined with \eqref{su_lap_end} gives
\begin{equation}\label{aquitard_flux}
\frac{\partial \bar{\bar{s}}}{\partial z}\bigg|_{z=0}=q_1(\bar{\bar{s}}_C+\bar{\bar{s}}_U)_{z=0}
\end{equation}
where $q_1=\frac{{{K}_{z1}}}{{{K}_{z}}}{{\eta }_{1}}\tanh \left( {{\eta }_{1}}{{b}_{1}} \right)$.

\section{Lapalce Transformed Unsaturated Zone Drawdown}
Lapalce transformed drawwdown  $\bar\sigma$ in the unsaturated zone are given by \citet{mishra_neuman2011} as
\begin{equation}
\bar{\sigma }(r_{D} ,z_{D,} p_{D} )=\left\{\begin{array}{l} {\int _{0}^{\infty }e^{a_{kD} \left(z_{D} -1\right)/2}
 \frac{\mathrm{J}_{n} [i\phi (z_{D} -1)]+\chi \mathrm{Y}_{n} [i\phi (z_{D} -1)]}{\mathrm{J}_{n} [i\phi (0)]+\chi \mathrm{Y}_{n} [i\phi (0)]} } \\ {{\rm \; }
\times \left(\bar{\bar{s}}_C+\bar{\bar{s}}_U \right)_{z_D=1}\frac{r_D^2K_D}{r^2}\;\mathrm{d}y{\rm\; \; \; \; \; \; \; \; \; \; for\; }a_{cD} \ne a_{kD} } \\
\\
{\int _{0}^{\infty }\frac{e^{\delta _{1D} \left(z_{D} -1\right)} +\chi e^{\delta _{2D} \left(z_{D} -1\right)} }{1{\rm +}\chi }  } \\ {{\rm \; }
\times \left(\bar{\bar{s}}_C +\bar{\bar{s}}_U \right)_{z_D=1}\frac{r_D^2K_D}{r^2} \;\mathrm{d}y {\rm\; \; \; \; \;  \; for\; }a_{cD} =a_{kD} =\kappa _{D} } \end{array}\right.
\end{equation}

where 
${{r}_{D}}=r/b$, 
${{z}_{D}}=z/b$, 
${{\mu}^{2}}={{y}^{2}}+
  \frac{{{p}_{D}}}{{{t}_{s}}{{K}_{D}}r_{D}^{2}}$, 
${{t}_{s}}={{\alpha }_{s}}t/{{r}^{2}}$, 
${{\alpha }_{s}}={{K}_{r}}/{{S}_{s}}$, 
${{q}_{D}}=qb$, 
${{a}_{kD}}={{a}_{k}}b$, 
${{a}_{cD}}={{a}_{c}}b$, 
$\phi ({{z}_{D}})=
  \sqrt{\frac{4{{B}_{D}}}{{{\lambda}_{D}}^{2}}}{{e}^{{{\lambda }_{D}}{{z}_{D}}/2}}$ , 
${{\lambda }_{D}}={{a}_{kD}}-{{a}_{cD}}$, 
${{B}_{D}}=
{{p}_{D}}\frac{{{S}_{D}}{{a}_{cD}}
  {{e}^{{{a}_{kD}}\left({{\psi }_{kD}}-{{\psi }_{aD}}\right)}}}{{{t}_{s}}{{K}_{D}}r_{D}^{2}}$, 
${{S}_{D}}={{S}_{y}}/S$,  
${{\psi }_{kD}}={{\psi }_{k}}/b$, 
${{\psi }_{aD}}={{\psi }_{a}}/b$, 
${{\delta }_{1D,2D}}=
{{\delta }_{1,2}}b=\frac{{{\kappa }_{D}}\mp
  \sqrt{{{\kappa }_{D}}^{2}+4\left( {{B}_{D}}+{{y}^{2}} \right)}}{2}$, 
$\nu =\sqrt{\frac{{{a}_{kD}}^{2}+4{{y}^{2}}}{{{\lambda }_{D}}^{2}}}$,
and 
${{p}_{D}}=pt$ are dimensionless quantities, $p$ being the Laplace
transform parameter;
\begin{equation}\label{wellbore_hank}
\begin{split}
{ \bar{\bar{s}}_C }(z=b)=
C_0\frac{{{r}^{2}}}{{{K}_{D}}r_{D}^{2}}\left\{ \frac{{{y}_{D}}{{r}_{wD}}}{{{y}^{2}}}{\mathrm{J}_{1}}\left( {{y}_{D}}{{r}_{wD}} \right){{\mathrm{K}}_{0}}\left( {{r}_{wD}}{{\phi }_{0}} \right)
+ \Gamma(0) \right\} {\rm \; \;\;\;\;}\\
+\sum\limits_{n=1}^{\infty }{{{C}_{n}}\frac{{{r}^{2}}}{{{K}_{D}}r_{D}^{2}}\left\{ \frac{{{y}_{D}}{{r}_{wD}}}{{{y}^{2}}}{\mathrm{J}_{1}}\left( {{y}_{D}}{{r}_{wD}} \right){\mathrm{K}_{0}}\left( {{r}_{wD}}{{\phi }_{0}} \right)
+ \Gamma(n) \right\}}
\end{split}
\end{equation}
where $\Gamma (n) = \frac{{{r}_{wD}}{{\phi }_{n}}{\mathrm{J}_{0}}\left( {{y}_{D}} \right){\mathrm{K}_{1}}\left( {{\phi }_{n}} \right)-{{y}_{D}}{{r}_{wD}}{\mathrm{J}_{1}}\left( {{y}_{D}} \right){\mathrm{K}_{0}}\left( {{\phi }_{n}} \right)}{{{\mu }^{2}}+{{n}^{2}}{{\pi }^{2}}}$, ${{y}_{D}}=yK_{D}^{1/2}{{r}_{D}}$, ${\mathrm{J}_{0}}$ and ${\mathrm{J}_{1}}$ being Bessel functions of first kind and, respectively, orders zero and one; and

\begin{eqnarray}
\chi =\left\{\begin{array}{ll}
-\frac{\left(a_{kD} +n\lambda _{D} \right)\mathrm{J}_{n} \left[i\phi \left(L_{D} \right)\right]-2i\sqrt{B_{D} e^{\lambda _{D} L_{D} } } \mathrm{J}_{n+1} \left[i\phi \left(L_{D} \right)\right]}{\left(a_{kD} +n\lambda _{D} \right)\mathrm{Y}_{n} \left[i\phi \left(L_{D} \right)\right]-2i\sqrt{B_{D} e^{\lambda _{D} L_{D} } } \mathrm{Y}_{n+1} \left[i\phi \left(L_{D} \right)\right]}  &a_{kD} \ne a_{cD}  \\
i & a_{kD} \ne a_{cD}, L_{D} \to \infty  \\
-\frac{\delta _{1D} }{\delta _{2D} } e^{\left(\delta _{1D} -\delta _{2D} \right)L_{D} } & a_{kD} =a_{cD}\\
0 & a_{kD} =a_{cD}, L_D \to \infty
\end{array}\right.
\end{eqnarray}

\begin{eqnarray}
q_D=\left\{\begin{array}{ll}
\left(\frac{a_{kD} }{2} +\frac{n\lambda _{D} }{2} \right)-i\sqrt{B_{D} } \frac{\mathrm{J}_{n+1} \left[i\phi \left(0\right)\right]+\chi \mathrm{Y}_{n+1} \left[i\phi \left(0\right)\right]}{\mathrm{J}_{n} \left[i\phi \left(0\right)\right]+\chi \mathrm{Y}_{n} \left[i\phi \left(0\right)\right]} & a_{kD} \ne a_{cD} \\
\frac{\delta _{1D} +\chi \delta _{2D} }{1+\chi } & {a_{kD} =a_{cD} =\kappa _{D} } \end{array}\right.
\end{eqnarray}
where $L_D = L/b$, $\mathrm{J}_n$ and $\mathrm{Y}_n$ being first and
second kind Bessel functions of order $n$.

\bibliographystyle{elsarticle-num-names}
\bibliography{leaky_unconfined}

\end{document}